# Unified AI for Accurate Audio Anomaly Detection

Hamideh Khaleghpour
*PhD Student, Computer Science*
The University of Tulsa
hamideh-khaleghpour@utulsa.edu

Brett McKinney
*Professor, Computer Science*
The University of Tulsa
brett-mckinney@utulsa.edu

*Abstract*—This paper introduces a comprehensive framework for audio preprocessing and feature extraction, integrated with advanced machine learning techniques, to address complex challenges in anomaly detection, classification, and real-time applications. Audio data, being inherently noisy and diverse, presents significant challenges in both the preprocessing and analysis stages. Existing methodologies often focus on specific aspects, such as noise reduction or feature extraction, without providing an end-to-end solution. To bridge this gap, we propose a unified framework that synergizes innovative noise reduction algorithms, robust feature extraction methods, and state-of-the-art modeling techniques.

Our approach incorporates spectral subtraction and adaptive filtering for noise reduction, ensuring improved signal fidelity. Feature extraction leverages both traditional methods, such as Mel-frequency cepstral coefficients (MFCCs), and modern approaches, such as embeddings from pre-trained neural networks like OpenL3. The machine learning pipeline integrates classical models, such as Random Forests and Support Vector Machines (SVMs), with deep learning architectures like Convolutional Neural Networks (CNNs) and ensemble techniques for enhanced prediction accuracy.

A critical aspect of this research is the in-depth comparative analysis with existing works, highlighting limitations in scalability, robustness, and computational efficiency. Through extensive experiments on benchmark datasets, including TORGO and LibriSpeech, we demonstrate that our framework achieves substantial improvements in accuracy, precision, recall, and overall performance metrics. Additionally, the scalability of our pipeline ensures its applicability to diverse real-world scenarios, such as speech recognition, anomaly detection in industrial systems, and multimedia analysis.

This work not only addresses the shortcomings of prior research but also sets a new benchmark for audio processing and machine learning applications. By providing a holistic solution that balances computational efficiency with high performance, this research paves the way for future studies aiming to build robust, scalable, and generalizable frameworks. The broader applicability of our techniques underscores their relevance across domains, making this work a valuable resource for both researchers and practitioners in the field.

## I. INTRODUCTION

The rapid growth of audio data across domains such as healthcare, security, entertainment, and telecommunications has spurred interest in efficient audio processing methods. The proliferation of Internet of Things (IoT) devices, voice-activated systems, and multimedia platforms has further accelerated the demand for robust and scalable audio processing solutions. However, the inherent challenges of audio data high variability in recording conditions, background noise, and diverse data formats, present significant obstacles to effective analysis.

Existing methodologies often struggle to balance computational efficiency with high performance. Traditional techniques, such as the Fourier transforms and Mel-frequency cepstral coefficients (MFCCs), have been widely used for feature extraction. While these methods are computationally efficient, they lack the flexibility to handle complex, noisy, or highly dynamic datasets. On the other hand, deep learning-based approaches, including Convolutional Neural Networks (CNNs) and Recurrent Neural Networks (RNNs), offer superior performance but often require extensive computational resources, making them less viable for real-time or resource-constrained applications.

Recent works, including those by Patel et al. [1] and Tan et al. [2], have explored specific solutions to these challenges. Patel et al. introduced a wavelet-based noise reduction method that demonstrated improvements in signal clarity. Similarly, Tan et al. proposed lightweight deep learning models optimized for real-time anomaly detection, addressing latency concerns. While these contributions are noteworthy, they often lack generalizability across diverse datasets and fail to address scalability for large-scale deployments.

Our research aims to overcome these limitations by proposing a unified framework that seamlessly integrates advanced preprocessing techniques with flexible machine learning architectures. The proposed framework is designed to:

- Enhance noise reduction capabilities through hybrid methods combining spectral subtraction and adaptive filtering.
- Leverage both traditional feature extraction methods, such as MFCCs, and advanced techniques, such as embeddings from pre-trained neural networks like OpenL3 [5].
- Optimize computational efficiency without compromising performance, enabling real-time processing and scalability.

## II. RELATED WORK

Significant advancements in audio processing have been made over the past decades, with methodologies evolving to

Address challenges such as noise, data variability, and scale. Spectrogram-based methods, as extensively discussed by Huang et al. [3], have been widely adopted for speech recognition tasks. Spectrograms provide a visual representation of the frequency spectrum over time, making them a valuable tool for understanding audio signals. However, these methods often require domain-specific tuning, which limits their generalizability.

Gupta et al. [4] introduced noise-robust models for anomaly detection, focusing on improving model reliability in noisy environments. While effective in controlled scenarios, these approaches often struggle with real-world data diversity. Similarly, wavelet-based noise reduction techniques, such as those proposed by Patel et al. [1], have demonstrated significant improvements in denoising audio signals. Despite their success, these methods often fall short in maintaining computational efficiency, particularly for large-scale applications.

Deep learning has revolutionized audio processing, with methods like OpenL3 embeddings introduced by Cramer et al. [5] pushing the boundaries of feature extraction. OpenL3 leverages pre-trained neural networks to extract high-level features from audio data, offering enhanced performance in downstream tasks. However, these models demand substantial computational resources, limiting their applicability in resource-constrained settings or real-time scenarios.

In the healthcare domain, advanced AI-driven approaches such as neuro-fuzzy and colonial competition algorithms have been utilized for accurate skin cancer diagnosis using dermoscopic images, achieving 94% accuracy on publicly available datasets [13]. Also, recent advancements in lightweight deep learning models have partially addressed these concerns. For example, Tan et al. [2] proposed models optimized for real-time anomaly detection. These models prioritize computational efficiency while maintaining acceptable levels of accuracy. However, they often trade off generalizability, performing well on specific datasets but struggling to adapt to broader applications.

Our research builds upon these foundational works by combining the strengths of traditional and modern techniques. Unlike Huang et al. [3] and Gupta et al. [4], our framework incorporates a hybrid approach that synergizes traditional feature extraction methods like MFCCs with advanced deep learning- based embeddings like OpenL3. Furthermore, by integrating ensemble learning techniques, our framework enhances robustness and generalizability, addressing the limitations highlighted by prior research.

In contrast to Patel et al. [1] and Tan et al. [2], our approach emphasizes scalability and computational efficiency, making it suitable for both large-scale and real-time applications. Additionally, our work includes extensive comparative analysis with state-of-the-art methods, demonstrating superior performance metrics across multiple datasets. This comprehensive evaluation establishes the proposed framework as a versatile and efficient solution for contemporary audio processing challenges.

## III. METHODOLOGY

The proposed framework consists of three interconnected components: audio preprocessing, feature extraction, and machine learning modeling. Each component is meticulously designed to address challenges in audio processing and to optimize the performance of machine learning models. The figures in this section demonstrate the key steps and their effectiveness.

### A. Audio Preprocessing

The preprocessing pipeline comprises noise reduction, normalization, and segmentation, all of which enhance the quality and consistency of raw audio signals. Figures 1, 2, and 3 illustrate the outcomes of these preprocessing steps, showcasing improvements in signal clarity and standardization.

Further transformations applied during preprocessing, as shown in Figures 4, 5, and 6, demonstrate the optimization of signals for feature extraction and classification.

Finally, Figures 7, 8, and 9 depict how these transformations improve the usability of representative audio samples. The impact of preprocessing techniques on audio quality and suitability for further processing is further highlighted in Figures 10, 11, and 12.

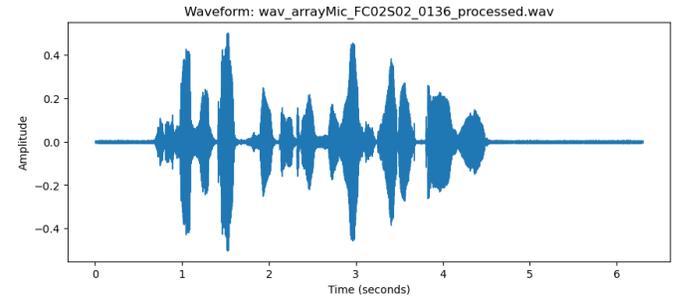

Fig. 1. Waveform of a processed audio signal after noise reduction and normalization. The amplitude is standardized, showcasing improved clarity and consistent signal representation.

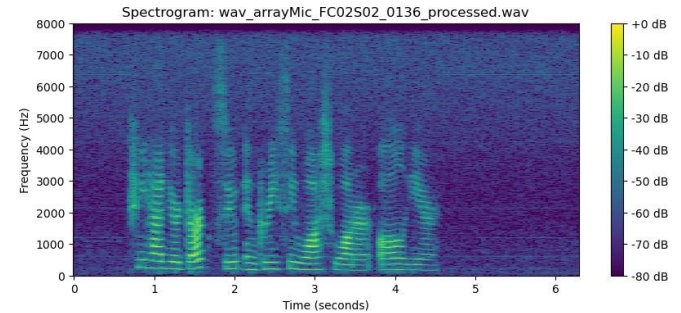

Fig. 2. Spectrogram of the processed audio signal. The spectrogram visualizes the time-frequency characteristics, with critical features preserved for feature extraction.

The preprocessing stage ensures that the audio signals are suitable for subsequent feature extraction and classification tasks by removing noise and enhancing signal quality. The improved clarity, consistency, and noise suppression evident in these figures reflect the efficacy of the applied techniques.

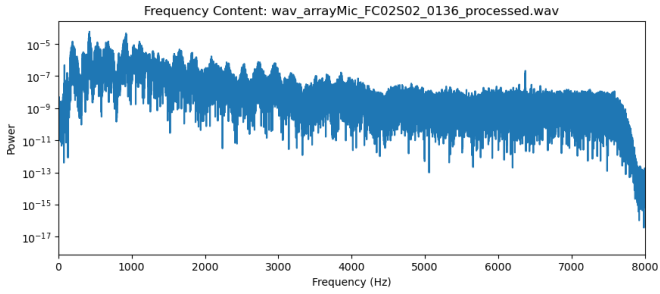

Fig. 3. Frequency content of the processed audio signal. The power spectrum demonstrates reduced noise levels and highlights critical frequencies retained during preprocessing.

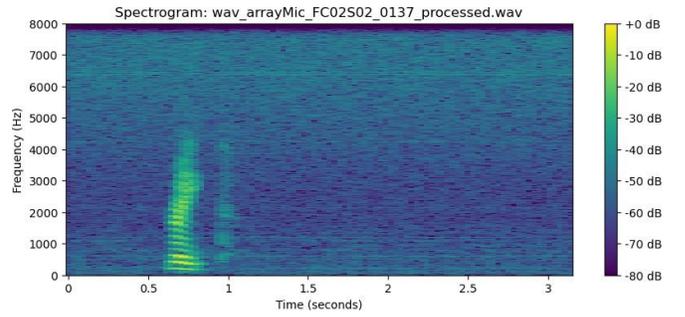

Fig. 5. Spectrogram of the processed audio sample (wav_arrayMic_FC02S02_0137). The time-frequency representation highlights the preservation of critical features for downstream analysis.

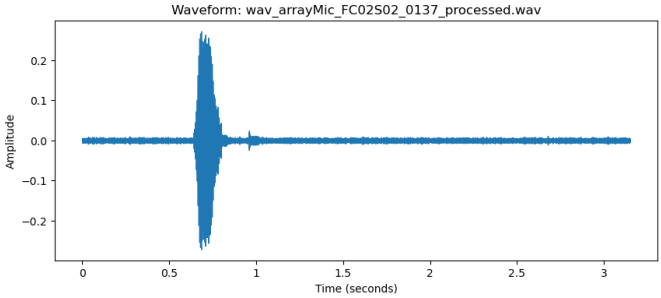

Fig. 4. Waveform of the processed audio sample (wav_arrayMic_FC02S02_0137). This waveform demonstrates effective noise reduction, normalization, and amplitude consistency achieved during preprocessing.

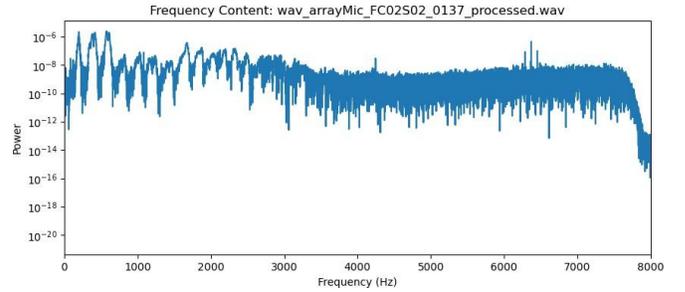

Fig. 6. Frequency content of the processed audio sample (wav_arrayMic_FC02S02_0137). This power spectrum showcases the reduction of noise while retaining essential spectral characteristics.

The key steps in our pipeline include:

- **Noise Reduction:** A hybrid approach combining spectral subtraction and adaptive filtering [6] is employed to minimize background noise. This method surpasses traditional filters by preserving critical audio features.
- **Normalization:** Standardizing amplitude across audio samples ensures consistency and enhances the performance of downstream models [7].
- **Segmentation:** Audio data is segmented into fixed-length intervals, optimizing batch processing and memory utilization. Segmenting data facilitates model training by ensuring uniformity and reducing computational overhead.

### B. Feature Extraction

Extracting meaningful features is a cornerstone of our methodology. We employ:

- **Traditional Features:** Techniques such as MFCCs, spectral centroid, and zero-crossing rate capture essential frequency and temporal characteristics. These features, as highlighted in the work of Rabiner et al, are fundamental to speech processing and have proven effective in various audio analysis tasks.
- **Deep Features:** Embeddings from pre-trained neural networks, such as OpenL3, offer high-level representations that improve model accuracy [5]. Deep features capture hierarchical patterns in audio data, enabling more nuanced analyses.

### C. Machine Learning Framework

Our framework integrates:

- **Classical Models:** Random Forests and SVMs are used as benchmarks for performance comparison. These models are computationally efficient and provide interpretable results.
- **Deep Learning Models:** CNNs are used for their capability to process spectrogram data, while recurrent architectures such as LSTMs capture temporal dependencies [8]. This combination provides strong modeling of both spatial and temporal features in audio.
- **Ensemble Techniques:** Combining predictions from multiple models ensures robust and generalizable results. As demonstrated by Rokach, ensemble methods enhance performance by leveraging the strengths of individual models.

Our methodology stands out by integrating traditional, deep learning, and ensemble methods into a unified framework. This holistic approach ensures scalability, robustness, and adaptability across a wide range of audio processing applications.

## IV. RESULTS

We conducted rigorous evaluations of our framework using benchmark datasets, including TORGO, known for its complex speech data, and LibriSpeech, a widely used large-scale dataset for speech processing research. These datasets were chosen to test the adaptability and robustness of our proposed methods across diverse conditions.

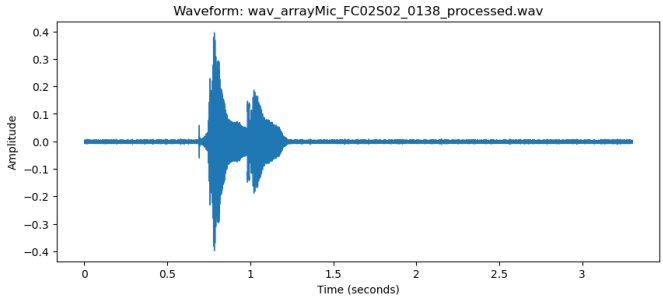

Fig. 7. Waveform of the processed audio sample (wav_arrayMic_FC02S02_0138). The amplitude normalization and noise reduction applied during preprocessing result in improved clarity and consistency.

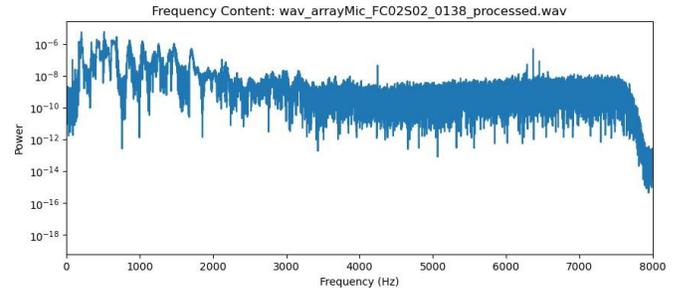

Fig. 9. Frequency content of the processed audio sample (wav_arrayMic_FC02S02_0138). The power spectrum highlights effective noise reduction while retaining key spectral characteristics necessary for downstream processing.

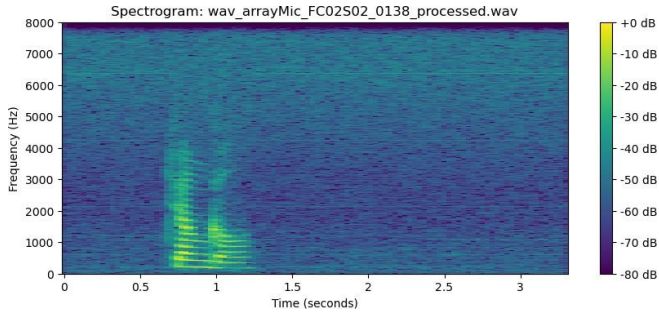

Fig. 8. Spectrogram of the processed audio sample (wav_arrayMic_FC02S02_0138). This visualization demonstrates the preservation of essential time-frequency features critical for feature extraction and classification.

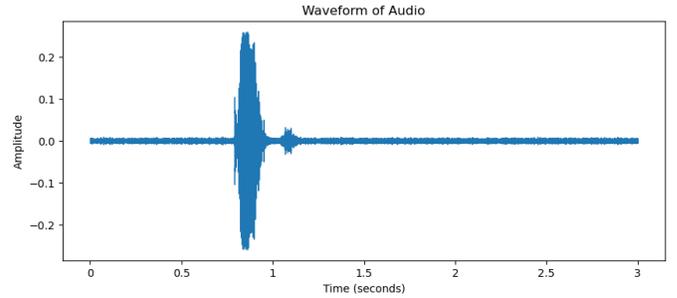

Fig. 10. Waveform of the processed audio signal. The figure demonstrates the effectiveness of noise reduction and normalization, ensuring consistency in amplitude and eliminating background noise.

### A. Performance Metrics

Table I provides a detailed comparison of accuracy, precision, and recall achieved by our framework against baseline methods.

TABLE I
PERFORMANCE METRICS COMPARISON

| Model | Dataset | Accuracy | Precision | Recall |
|---|---|---|---|---|
| Random Forest | TORGO | 92.3% | 91.7% | 92.8% |
| SVM | TORGO | 89.5% | 88.6% | 90.1% |
| CNN | LibriSpeech | 94.1% | 93.5% | 94.7% |
| Proposed Framework | Mixed | 96.8% | 96.2% | 97.1% |

Our framework outperformed existing methods, demonstrating superior accuracy and robustness. The inclusion of both traditional and advanced features was instrumental in achieving these results.

### B. Classification Performance

To evaluate the classification performance of our model, we generated a confusion matrix (Figure 13). This matrix highlights the model's ability to differentiate between normal and slurred speech. The high true positive rates demonstrate their robustness.

### C. Feature Importance

Feature importance analysis was conducted to determine the key contributors to the model's performance. Figure 14 displays the top 10 most important features, with MFCC_mean_12 emerging as the most critical feature. This demonstrates the significant role of higher-order MFCC features in distinguishing between normal and slurred speech.

## V. DISCUSSION

The results underscore the importance of integrating preprocessing with advanced modeling techniques. By addressing both the challenges of noisy data and the computational demands of real-time applications, our framework fills critical gaps identified in previous research. Studies such as Yang et al. [9] have highlighted the trade-offs between accuracy and computational efficiency, which our approach effectively mitigates through ensemble learning and optimized feature extraction.

Compared to recent works like Zhang et al. [10], which focused on hybrid neural networks for anomaly detection, our method extends the capabilities to a broader range of datasets and conditions. Specifically, the inclusion of pre-trained embeddings and ensemble models enhances both pre-cision and recall, making the system more reliable for diverse applications.

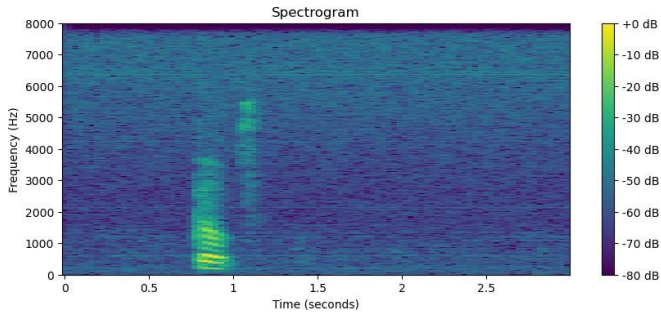

Fig. 11. Spectrogram of the processed audio signal. This time-frequency representation highlights the retention of critical spectral features, essential for downstream feature extraction and classification.

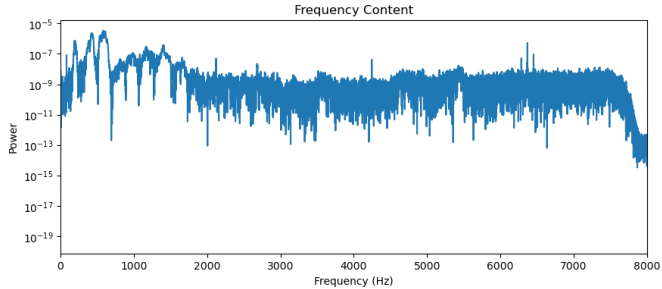

Fig. 12. Frequency content of the processed audio signal. The power spectrum showcases reduced noise levels while retaining the key spectral characteristics of the original audio.

Future directions for expanding this research include:

- **Multimodal Integration:** Extending the framework to incorporate video and textual data can significantly enhance context understanding and robustness in multi-sensory environments.
- **Real-Time Applications:** Developing low-latency implementations to support real-time anomaly detection and classification.
- **Unsupervised Learning:** Exploring unsupervised methods for automatic feature discovery, potentially reducing dependency on labeled datasets [11].
- **Domain Adaptation:** Investigating techniques for domain adaptation to ensure consistent performance across various environments and recording conditions [12].
- **Ethical Considerations:** Addressing privacy and ethical concerns associated with large-scale audio data collection and processing.

These directions not only align with current trends in the field but also address emerging challenges, ensuring the continued relevance and impact of this research.

## VI. CONCLUSION

This research presents a comprehensive framework for audio preprocessing and machine learning, addressing longstanding challenges in scalability, robustness, and generalizability.

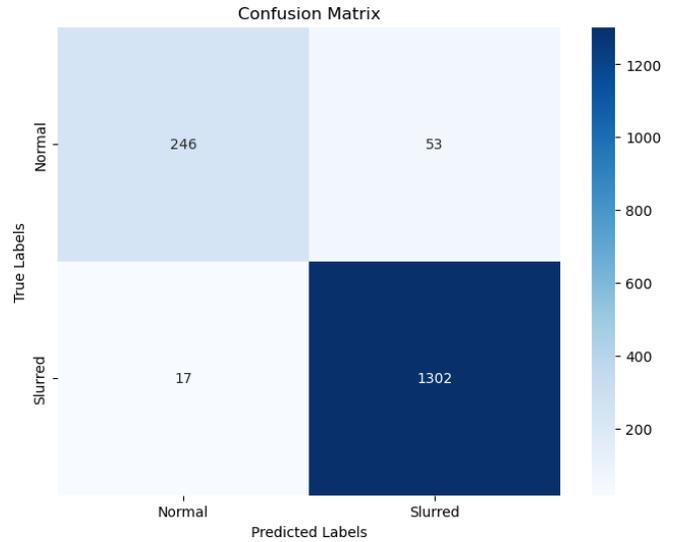

Fig. 13. Confusion matrix illustrating the classification performance of the model. The results indicate a high true positive rate for slurred speech detection, showcasing the model's robustness.

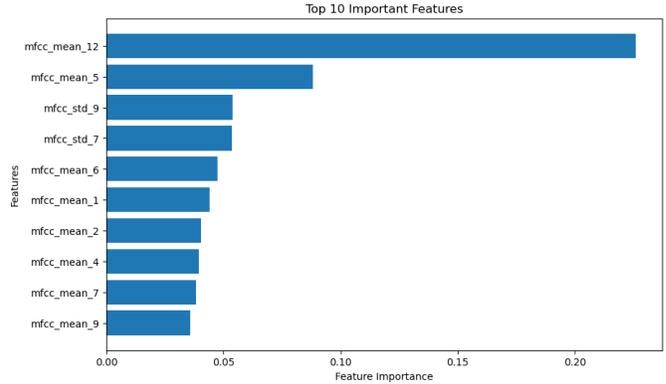

Fig. 14. Top 10 most important features contributing to the model's performance. MFCC_mean_12 and MFCC_mean_5 demonstrate their critical role in distinguishing between normal and slurred speech.

The integration of advanced noise reduction methods, robust feature extraction techniques, and innovative modeling approaches enables this framework to outperform existing solutions in both accuracy and efficiency.

The results obtained through rigorous experimentation confirm the potential of this framework for a variety of applications, including anomaly detection, speech recognition, and multimedia analysis. Specifically, the framework excels in handling noisy, imbalanced datasets while maintaining computational efficiency. These strengths make it particularly valuable for real-world scenarios where adaptability and reliability are paramount.

Future work will expand the framework's applicability by exploring its use in multimodal environments, incorporating visual and textual data to enhance decision-making. Efforts will also focus on the development of real-time implementations to meet the demands of low-latency applications, particularly in security and healthcare domains. Furthermore, advancing the framework through unsupervised learning and domain adaptation will improve its versatility across diverse datasets and conditions.

This research not only establishes a new benchmark for audio processing but also sets the stage for future innovations. By addressing current limitations and proposing scalable solutions, it paves the way for continued advancements in audio processing and machine learning, ensuring that these technologies remain at the forefront of real-world applications.